\documentclass[aps,prl,reprint,superscriptaddress]{revtex4-1}

\usepackage{amsmath,bm}
\usepackage[dvips]{graphicx,color}
\usepackage[latin1]{inputenc}
\usepackage[T1]{fontenc}
\setcitestyle{super}
\usepackage{latexsym}
\usepackage{times}
\usepackage{float}
\usepackage{soul}
\usepackage{url}

\begin{document}

\title{Demonstrating an absolute quantum advantage in direct absorption measurement}

\author{Paul-Antoine Moreau}
\email[Corresponding author: ]{paul.antoine.moreau@gmail.com}
\author{Javier Sabines-Chesterking}
\author{Rebecca Whittaker}
\affiliation{Quantum Engineering Technology Labs, H. H. Wills Physics Laboratory and Department of Electrical and Electronic Engineering, University of Bristol, Merchant Venturers Building, Woodland Road, Bristol BS8 1FD, UK}
\author{Siddarth K. Joshi}
\affiliation{Quantum Engineering Technology Labs, H. H. Wills Physics Laboratory and Department of Electrical and Electronic Engineering, University of Bristol, Merchant Venturers Building, Woodland Road, Bristol BS8 1FD, UK}
\affiliation{Now at: Institute for Quantum Optics and Quantum Information (IQOQI) Austrian Academy of Sciences, Boltzmanngasse 3, A-1090 Vienna, Austria}
\author{Patrick Birchall}
\author{Alex McMillan}
\author{John G. Rarity}
\affiliation{Quantum Engineering Technology Labs, H. H. Wills Physics Laboratory and Department of Electrical and Electronic Engineering, University of Bristol, Merchant Venturers Building, Woodland Road, Bristol BS8 1FD, UK}
\author{Jonathan C. F. Matthews}
\email[Corresponding author: ]{jonathan.matthews@bristol.ac.uk}
\affiliation{Quantum Engineering Technology Labs, H. H. Wills Physics Laboratory and Department of Electrical and Electronic Engineering, University of Bristol, Merchant Venturers Building, Woodland Road, Bristol BS8 1FD, UK}

\date{\today}

\begin{abstract}
\noindent 
Engineering apparatus that harness quantum theory offers practical advantages over current technology. A fundamentally more powerful prospect is the long-standing prediction that such quantum technologies could out-perform any future iteration of their classical counterparts, no matter how well the attributes of those classical strategies can be improved. Here, we experimentally demonstrate such an instance of \textit{absolute} advantage per photon probe in the precision of optical direct absorption measurement. We use correlated intensity measurements of spontaneous parametric downconversion using a commercially available air-cooled CCD, a new estimator for data analysis and a high heralding efficiency photon-pair source. We show this enables improvement in the precision of measurement, per photon probe, beyond what is achievable with an ideal coherent state (a perfect laser) detected with $100\%$ efficient and noiseless detection. We see this absolute improvement for up to $50\%$ absorption, with a maximum observed factor of improvement of 1.46. This equates to around $32\%$ reduction in the total number of photons traversing an optical sample, compared to any future direct optical absorption measurement using classical light. 
\end{abstract}

\maketitle

How to maximise information gained from each photon used to probe an unknown sample is a central question for metrology. It also has practical implications whenever we face limits on the brightness of optical probes---for example, where samples are altered or damaged by light, it is highly desirable to maximise information per unit of exposure~\cite{TaylorBowen2016}. By generating various quantum states of light, it is being demonstrated that one can improve upon the sensitivity and precision of current classical measurement techniques~\cite{he-prl-59-2555,mertz1990observation,ribeiro1997sub,hayat1999reduction,giovannetti2004quantum,d2006transmittivity,bondani2007sub,kalachev2007biphoton,blanchet2008measurement,BrambillaCaspaniJedrkiewiczEtAl2008,brida2010experimental,WolfgrammVitelliBeduiniEtAl2013,TaylorJanousekDariaEtAl2013,TaylorBowen2016}.  
An approach that allows sub-shot-noise-limit (SNL) measurement of an unknown sample's transmission is to use correlated beams of photons~\cite{JakemanRarity1986,AdessoDellAnnoDeSienaEtAl2009,he-prl-59-2555,TapsterSewardRarity1991,brida2010experimental}. But demonstrating unambiguous real-world enhancements using correlated photons has remained a challenge~\cite{thomas2011real}, and improvements in the efficiency of commercial low intensity CCD cameras in recent years to $>95\%$~\cite{2016} now means that previously reported quantum correlated measurements can be exceeded by direct measurement with classical light. 

\begin{figure}[b]
	\centering
		\includegraphics[width=\columnwidth]{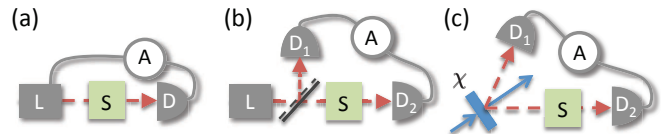}
	\caption{\textbf{Absorption measurement strategies.} \textbf{(a)} Direct absorption measurement with a single beam. \textbf{(b)} Differential measurement to suppress experimental classical noise. \textbf{(c)} Use of correlated twin beams generated from a nonlinear process ($\chi$). In each cartoon, D denotes detectors, S denotes a sample being measured and A denotes analysis that combines the intensity measurement of one detector with another or with a known mean intensity of a laser (denoted L).}
	\label{CartoonFig}
\end{figure}

Intensity measurement of an idealised laser fluctuates with a Poisson distribution~\cite{monras2007optimal,escher2011general,brida2010experimental}. This is typically used to define the SNL in optical measurements---which defines the limit of precision in optically measuring experimental parameters---and it can only be reached in classical experiments once all other sources of noise are removed. Accounting for information per-photon exposure (PPE) is particularly important for benchmarking schemes considered for measuring photo-sensitive samples~\cite{TaylorBowen2016}. For measuring optical transmission, the proportion of photons that pass through a sample is used to directly estimate absorption $\alpha$, and so the minimum error in estimation, $\Delta^2 \alpha$, is defined by the SNL. If we then fail to detect a subset of these photons---due to poor detection efficiency or component loss---exposure must be increased to compensate, which in turn reduces the information gained PPE.
The reduction of precision PPE can also manifest to varying degrees in different strategies---for example, noise introduced by reduced detector efficiency can have a greater impact on schemes using two detectors than on schemes designed to just use one detector (Fig.~\ref{CartoonFig}).
It is also important to note that for measurement schemes where the precision itself varies with parameters of the measured sample it is possible for precision to be reduced, potentially requiring either prior information about the optical sample or the addition of iterative feedback and control to ensure sub-shot-noise perfomance~\cite{Y.:2011aa,Yonezawa1514,Berni:2015aa}. For example, the use of amplitude squeezed light to measure reduced intensity fluctuations with homodyne detection~\cite{xiao1988detection,schneider1996bright} means precision is also dependent on optical phase introduced by the object being measured, leading to a random amount of precision that can be above the SNL.

\begin{figure*}
	\centering
		\includegraphics[width=\textwidth]{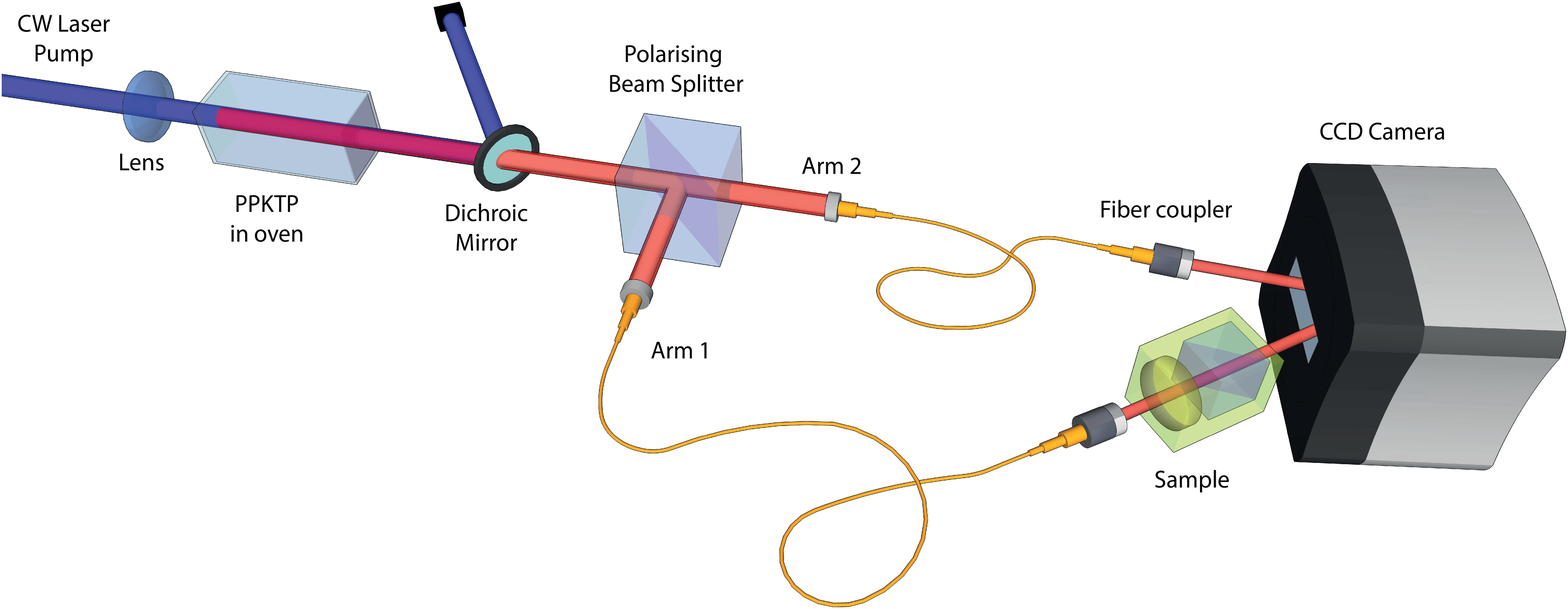}
	\caption{\textbf{Experimental setup.} A 404-nm laser beam pumps a PPKTP crystal to generate SPDC. After removing the pump beam, the two photons (each labeled the signal and idler photons) are separated and launched inside two optical fibres. The output of one fibre is passed through a sample before being focussed onto a few pixels of a CCD camera, while the second fibre output is directly focussed on the camera.}
	\label{Setup}
\end{figure*}

The reduced noise of quantum-correlated twin-beams generated with spontaneous parametric downconversion (SPDC) of a laser can be measured by recording correlated intensity using photodiodes~\cite{he-prl-59-2555,TapsterSewardRarity1991,iskhakov2016heralded}. 
For practical application, this is attractive because correlations between photon pairs are unaffected by optical phase induced by a measured object. This regime of detection can be preferable to photon counting because it reduces the amount of recorded data that is associated with the measurement of individual photons in the macroscopic states that are needed for most imaging and sensing scenarios. This technique can be transferred to detecting correlated photons altogether in the same image of a CCD camera, to image sub-SNL noise fluctuation of spatial correlations from SPDC~\cite{BrambillaCaspaniJedrkiewiczEtAl2008}. With the inclusion of a spatially absorbing sample, it has been shown that SPDC can be used to suppress noise in imaging objects to a degree that out-performs classical laser-measurement using an equally efficient detection~\cite{brida2010experimental}.
However, the performance reported in Ref.~\onlinecite{brida2010experimental} can now be exceeded by direct optical absorption measurement with classical light by $8\%$ PPE using currently available CCD camera efficiency of $95\%$~\cite{2016}. To extend such results to the claim that quantum correlated light can out perform any classical version of this measurement, it would need to be assumed that the undetected photons, were they detected, would give more information about the sample than the information gained from detected photons. Removing such an assumption is analogous to removing a fair sampling assumption in tests of Bell non-locality~\cite{peter2014testing,br-rmp-86-419}.

Here, we report precision PPE in measuring absorption that is up to $45\%$ beyond what is achievable with ideal direct optical absorption measurement using classical light. We do this using a $90\%$ efficient commercial air-cooled CCD, a new unbiased estimator for data analysis that maximises precision gained in our experiment and a high heralding efficiency photon-pair source enabled by recent developments of efficient sources of entangled pairs of photons~\cite{JOSHI2014}. This enables observation of sub-SNL physics with quantum correlated light without assuming fair sampling. It is also a direct comparison of current quantum experiments to the SNL of any future classical implementation directly illuminated by idealised Poisson distributed light and recorded with 100\% efficient noiseless detection. 
The absolute sub-shot-noise performance that we observe with a relatively simple experimental setup makes future use of quantum correlated light for enhanced measurement practical for applications.

\textbf{Experimental Setup:} Our experimental setup is presented in Fig.~\ref{Setup}: a periodically poled potassium titanyl phosphate (PPKTP) crystal designed for type-II phase matching is pumped with a continuous wave laser beam at $404$ nm to generate photon pairs---the signal photons emitted in arm 1 have wavelength $798$ nm and the idler photons emitted in arm 2 have wavelength $818$ nm, which is stabilised with the crystal temperature set to $24^\circ \text{C}$. A dichroic mirror then reflects the pump and transmits the down-converted photon pairs. Type-II phase matching ensures the signal and idler photons in each emitted photon pair are vertically and horizontally polarised respectively, and they are separated with a polarisation beam-splitter (PBS) before being injected into two single mode optical fibres. The whole photon pair generation setup is compact ($<$ 20 cm x 60 cm) and it is designed for high-efficiency collection into optical fibre: $\sim69$\% of the photons leaving the fibres are paired. The output of the two fibres are then imaged on a CCD camera, an Andor iDus 416 cooled to $-35^\circ \text{C}$, originally developed for spectroscopy. The camera exhibits a device detection efficiency of $\eta_d = 90\pm3\%$ under these conditions, this value has been experimentally characterised using calibrated detectors and is a lower bound according to manufacturer data. The varied absorption of a `sample' is implemented by a PBS preceded by a rotatable half-wave plate (HWP) and is intercalated between one of the fibres and the camera on the path of the beam (arm 1).\\

\begin{figure}
	\centering
	\hspace*{-4mm}
		\includegraphics[width=0.53\textwidth]{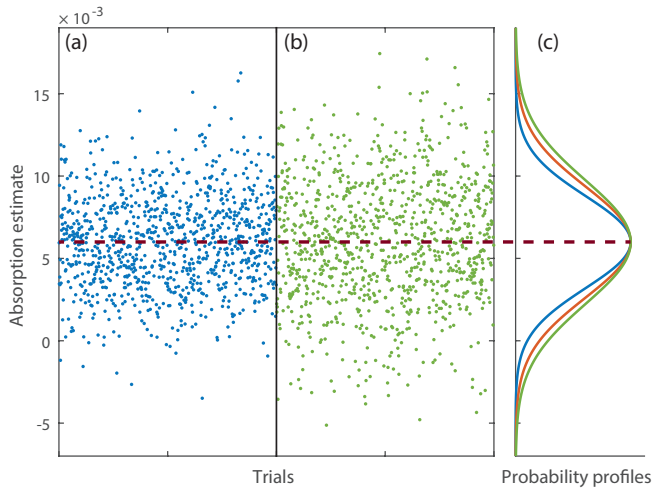}
	\vspace{-1cm}
	\caption{\textbf{Illustrating the improvement achieved using two-photons compared to single photons in practice.} (colour online) \textbf{(a)} Data points each obtained with a single acquisition of correlated two-photon states captured by the CCD and analysed with our estimator eq.~(\ref{estimator}). \textbf{(b)} Data points each obtained with a single acquisition of single-photon states and with the direct and sub-optimal estimator eq.~(\ref{estims}). Both data in (a) and (b) are each from 1,000 individual absorption estimates ($\alpha_s = 5.99\cdot 10^{-3}$ which is obtained with only the HWP as a sample). \textbf{(c)} The distributions of these data are compared by fitting Gaussian functions---the narrower blue curve corresponds to (a), the wider green curve corresponds to (b). The intermediate red curve corresponds to the expected distribution of estimates that would occur with a perfect coherent state measured with 100\% efficient and noiseless detection. The three distributions are presented with different normalisations so that their maximum are superposed along the horizontal axis to highlight the difference in width.}
	\label{Cl}
\end{figure}

\begin{figure*}
	\centering
		\includegraphics[width=0.5\textwidth]{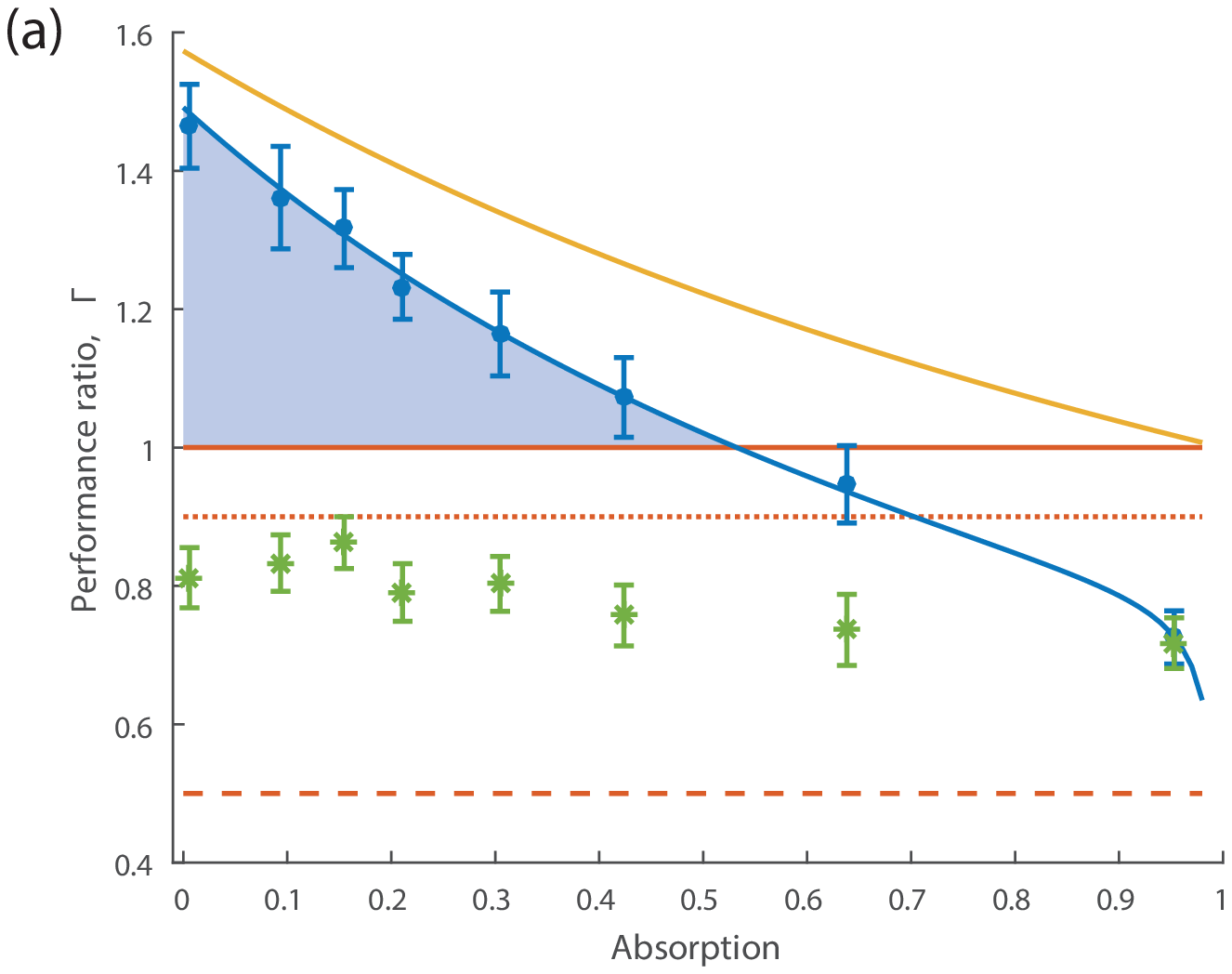}\hfill
		\includegraphics[width=0.5\textwidth]{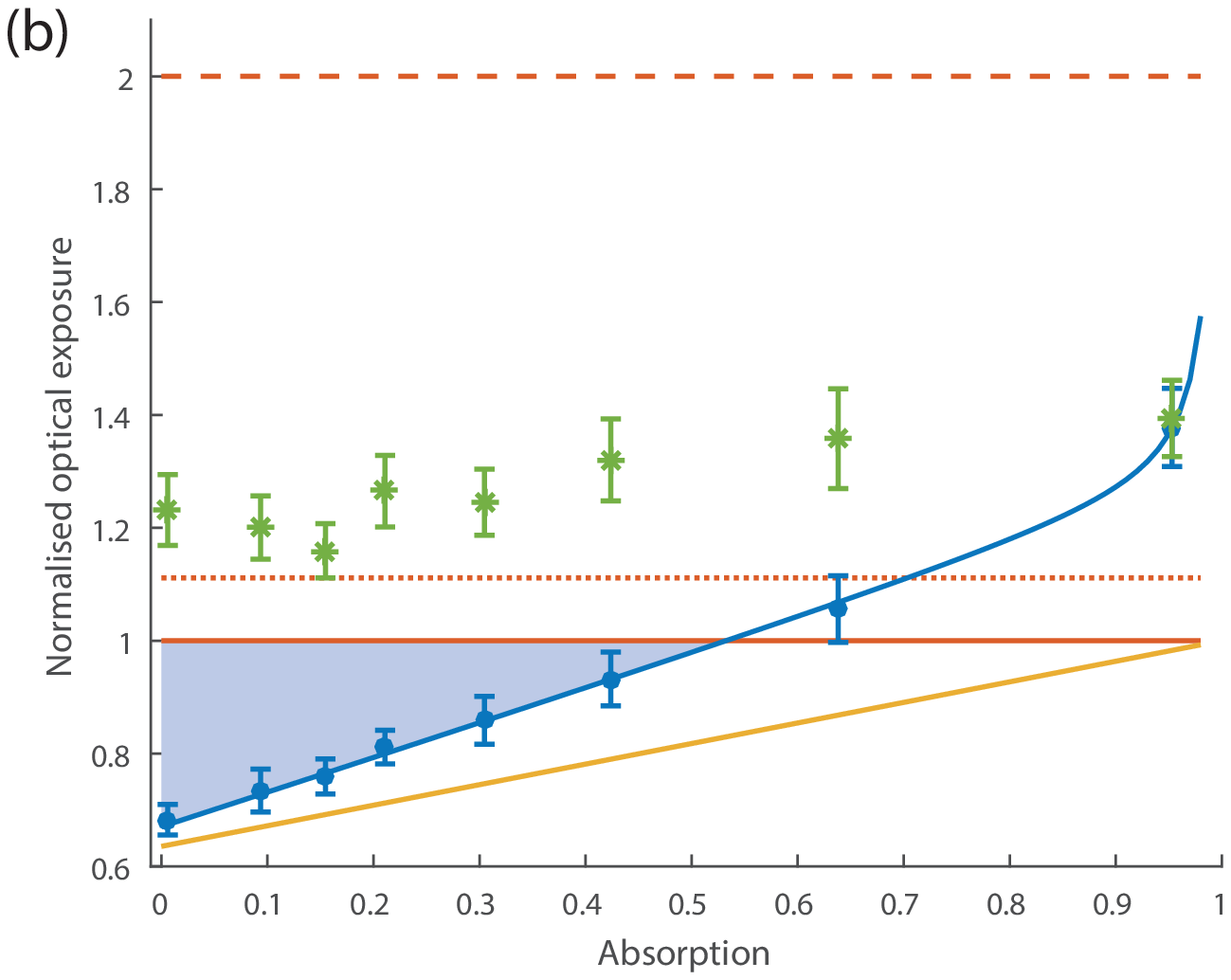}
	\caption{\textbf{Absolute quantum advantage in absorption measurement.} \textbf{(a)} $\Gamma$ is the ratio of the variance of an absorption estimate for the ideal classical scheme, to our experimental measurement. \textbf{(b)} Normalised optical exposure for equal precision, computed from the ratio between the number of photons probing the sample in a particular scheme and in the best direct classical scheme attaining the same estimation precision by increasing the illumination. In both plots: the red solid lines correspond to the ideal classical limit, the red dotted lines correspond to a classical state detected with the same $90\%$ efficiency as in our setup, and the red dashed lines correspond to the example of a differential classical measurement where a laser is split on a 50:50 beamsplitter (Fig.~\ref{CartoonFig} (b)) and therefore quantum fluctuations are un-correlated. The yellow line corresponds to the theoretical limit achievable by our system, taking into account the measured rate of classical fluctuations and the arms efficiencies. The blue dots with error bars correspond to the experimental data using the quantum corrected measurement. The green asterisks correspond a classical measurement with our setup, using only the single photons passing through the sample in Arm 1 and ignoring any correlated data in Arm 2. And finally the blue shadowed area highlight the absolute quantum advantage that is detected. Error bars correspond to standard deviations on the related quantities taking into account both the uncertainty on the precision of the measurement and the uncertainty on the efficiency of the camera. The blue lines correspond to best-fit to the theory, with extra parameters corresponding to eventual super-poissonian non-deterministic and deterministic noise in each fluorescence beam, and to camera noise. All data and curves in both plots normalise to PPE to the sample. N.b. the first point of each plot is obtained with only the HWP as a sample i.e. removing the PBS.}
	\label{QA}
\end{figure*}

\textbf{Theory:} When using one optical beam, absorption can be found using the standard estimator
\begin{equation}
\alpha_{s}=1-\eta_{s}=1-\frac{N_1}{\left\langle N_{1p}\right\rangle},
\label{estims}
\end{equation}
where $\eta_{s}$ is the transmission efficiency of the sample, $N_1$ is the number of photons detected on arm 1 with the sample and $N_{1p}$ is the mean number of photons detected on arm 1 without the sample. After each experimental trial, we use an estimator that gives an estimate of $\alpha_{s}$ that is ideally close to $\left\langle \alpha_{s}\right\rangle$ for each trial---the closer $\alpha_s$ is to its actual mean value $\left\langle \alpha_{s}\right\rangle$, the smaller the error on estimating absorption. As eq.~(\ref{estims}) shows, this estimation is limited by the evaluation of $N_1$ which fluctuates according to the quantum nature of light and manifests as noise when estimating $\langle N_1\rangle$.

Photon pairs emitted from two-mode SPDC fluctuate in number, but the fluctuation in total number of photons in each of the two output modes, $N_1$ and $N_2$, are correlated~\cite{he-prl-59-2555,TapsterSewardRarity1991}.
Noise on measuring $N_1$ can therefore be suppressed using the knowledge gained by measuring $N_2$. For example, if for a single trial $N_2$ is measured to be more than its expectation value $\left\langle N_2\right\rangle$, we know that the intensity $N_{1}$ will have exhibited the same fluctuation simultaneously and therefore that this value is also greater than its expectation value $\left\langle N_1\right\rangle$.
These sub-Poissonnian fluctuations can be used to improve the estimation of the number of photons $N_1$ after the sample therefore decrease statistical error in measuring optical absorption~\cite{TapsterSewardRarity1991}. Due to loss in an optical measurement however, uncorrelated single photons are also present in the experiment. Therefore, to maximise information available in our experiment, we have developed a new and unbiased estimator that  uses all detected photons that pass through a sample to estimate its absorption with quantum-enhanced noise-suppression PPE. This estimator is given by
\begin{equation}
\alpha_{s}=1-\frac{N'_1+\delta E}{\left\langle N_{1p}\right\rangle}=1-\frac{N_1-k\Delta N_2+\delta E}{\left\langle N_{1p}\right\rangle},
\label{estimator}
\end{equation}
where $N'_1$ is a corrected estimate for $N_1$ that is a function of: $\Delta N_2$, the deviation of $N_2$ from its mean value $\left\langle N_{2}\right\rangle$; $\delta E$, a small correction used to remove bias in the estimator, calibrated by taking images without a sample; and $k = CN_1$, a correction factor with constant $C$ that encapsulates the sources of noise in our photon source obtained during the calibration phase. In contrast to previous demonstrations, the new estimator eq.~(\ref{estimator}) is designed to use all detected light that has passed through the sample---both the correlated light that provides access as Fock state behaviour \cite{Javierinprep} and the classical contribution due to photons arriving on the sample with no detected twin due to system loss. Note that this strategy does not need to measure the arrival of individual photons, it only needs to record the fluctuation of large photon number. Therefore the main requirement for detection is high quantum efficiency.

In contrast, the performance of a perfect direct classical scheme is given~\cite{WhittakerErvenNevilleEtAl2015} by
\begin{equation}
\Delta^2 \alpha_{cl}=\frac{(1-\alpha)\eta_d}{\left\langle N_{1p}\right\rangle},
\label{varcoh}
\end{equation}
where we have increased the average detected intensity in the perfect classical model by a factor of $1/{\eta_d}$ in order to make a fair comparison with our experiment that uses detection efficiency $\eta_d$.
To quantify a quantum advantage, we compute a performance parameter $\Gamma$ that is a ratio of the minimum noise in a classical measurement $\Delta^2 \alpha_{cl}$ to the measured variance of absorption estimates
\begin{equation}
\Gamma=\frac{\Delta^2 \alpha_{cl}}{\Delta^2 \alpha_{\text{exp}}}.
\label{ratio}
\end{equation}
$\Gamma>1$ signifies a quantum advantage in precision PPE.\\

\textbf{Results:} To characterise our setup, we recorded vertically binned images using the setup presented on Fig. \ref{Setup} without the PBS and HWP implementing the sample. This enables characterisation of $\left\langle N_2\right\rangle$, $\left\langle N_{1p}\right\rangle$, $k$, $\delta E$.
We find the normalised variance of the difference between the two beam intensities to be $\sigma=\frac{\Delta^2(N_1-N_2)}{N_1+N_2}=0.38\pm0.02$, which is a witness of strong sub-shot noise statistics since $\sigma=0$ corresponds to fully correlated statistics and $\sigma=1$ corresponds to two perfectly independent beams fluctuating at the SNL. For comparison to single photon counting experiments, this corresponds to a Klyshko heralding efficiency of {$62\%$}. 

Although the evaluation of the absorption could be done with just one acquisition of $\sim$0.5s, in practice we measured 1000 acquisitions (10 series of 100 acquisitions) to estimate $\Delta^2\alpha_{\text{exp}}$ which corresponds to a total acquisition time of 15 minutes. We repeated this process for each measured value of absorption. We corrected the value of $\left\langle N_{1p}\right\rangle$ for each set of 100 acquisitions using the mean value $\left\langle N_2\right\rangle$ obtained during those same acquisitions to avoid bias in the estimation of the absorbance due to the classical fluctuations of the source. For each image acquisition we make an estimation of $\alpha_s$ using the estimator in eq.~(\ref{estimator}) as presented in Fig. \ref{Cl}, we use the 100 acquisitions of the series to evaluate $\Delta^2 \alpha_{\text{exp}}$ and reproduce this for 10 series in order to evaluate the experimental uncertainty on the mean value of $\Delta^2 \alpha_{\text{exp}}$. We can see in Fig. \ref{Cl} (c), the mean absorption of the probability profiles of the blue and green curves overlap, indicating that our estimator eq.~(\ref{estimator}) is unbiased.

As shown in Fig. \ref{Cl} with the example of measuring $\alpha_{s}=0.04$, our estimator eq.~(\ref{estimator}) (shown in panel (a)) gives a reduced dispersion of the estimation compared to the best direct classical scheme for the same number of photons probing the sample, marked by the red theoretical curve in panel (c). We measure $\Delta^2 \alpha_{\text{exp}}$ over the range $0<\alpha_s<1$ and compare in Fig. \ref{QA}(a) to the ideal performance of perfect classical direct detection using the precision ratio $\Gamma$ (eq.~(\ref{ratio})). For $0<\alpha_s<0.5$ we observe $\Gamma>1$, signifying an absolute quantum advantage. The highest advantage that we see is a reduction in estimate variance by $46\pm6\%$ for $\alpha_s = 5.99\cdot 10^{-3}$, which is distinguished by more than 7 standard deviations from the classical limit. This also equates to an advantage of $63\pm7\%$  over a direct classical scheme measured with $90\%$ detection efficiency, that corresponds to the efficiency of the CCD camera used in our experiment. Since for both the reported strategy and the perfect classical strategy, the variance scales linearly with $N_p$, it follows for an equal precision that the ratio between the total number of probe photons used in the reported quantum measurement and the perfect classical strategy is given by $1/\Gamma$. Fig. \ref{QA}(b) shows this reduced number of probe photons using our experimental data compared to a perfect classical scheme achieving the same measurement precision. The highest effective reduction of probe photons we observe is $32\pm3\%$ ($1.7$ dB), which is also a reduction by $39\pm3\%$ ($2.1$ dB) compared to a classical method using $90\%$ detection efficiency.

\textbf{Discussion:} To conclude, we have demonstrated an absolute quantum advantage over the best classical optical strategy for direct absorption measurement over the range $0<\alpha<0.5$. This demonstrates that quantum-enhanced absorption measurement need not be confined to measuring weak absorption. From a practical perspective, the scheme we have used is not limited by requirements of high temporal resolution of coincidence detection, data storage or equivalently low photon flux requirements associated to individual photon counting. Therefore, future experiments can be performed with high photon flux such as from high-brightness correlated photon sources~\cite{mertz1990observation}. In fact recent experiments using relatively bright (pulsed) SPDC light~\cite{iskhakov2016heralded} suggest a noise reduction factor up to $83.4\%$ ($\sigma=0.166$) could be achieved if detector noise and classical fluctuations can be suppressed. Similarly, by further optimising our collection efficiency,  we expect to be able to report much higher precision and greater noise reductions in future. Furthermore the technique reported here is of immediate use to obtain absolute quantum advantages in imaging~\cite{brida2010experimental} and absorption spectroscopy~\cite{WhittakerErvenNevilleEtAl2015}, in particular when used with narrow linewidth photon pair generation~\cite{mertz1990observation,WolfgrammVitelliBeduiniEtAl2013}. The relative simplicity of the scheme can also motivate analogous setups applied to high-energy optics---for example, the demonstrations of X-ray parametric down-conversion~\cite{Eisenberger1971,Shwartz2012} and the development of high efficiency direct hard X-ray detector arrays~\cite{Meuris2008} could be used to minimise radiation exposure in medical imaging.

\textbf{Method:} 
The photon pair source has been designed to optimise the collection of the spontaneous-down converted photons after their emission in the crystal. First, focusing conditions were chosen to maximise mode matching between the down-conversion modes and the collection modes using both numerical calculation to determine the initial design and improved experimentally by beam profiling. The mode matching has been estimated to be $80\%$. Another critical point is efficient removal of the pump beam by using a combination of a dichroic mirror and a colour glass filter. The collection of the twin photons in mono-mode fibres also acts as spatial filtering for parasitic light from the UV laser pump, that unlike the SPDC are not mode matched with the fibre modes. The transmission of the SPDC photons on each filter is $98\%$. To avoid fluorescence of the collection lenses due to the UV laser pump, we used Fused Silica optics. All the optics, including the PPKTP crystal, have anti-reflection coating with less than $0.2\%$ loss per surface. The fibres were coated only on one end and so add an extra $4\%$ loss. We can therefore predict a heralding efficiency not taking into account the detector efficiency of about $72\%$ without detector efficiency and $65\%$ including the $90\%$ detector efficiency. This is compatible with the experimental results giving a $62\%$ heralding efficiency with the camera.

We characterised the photon source using Excelitas Avalanche Photo-Diodes (APD) Model SPCM-800-14 that have a detection efficiency as certified by the manufacturer of $>62\%$. We measured an overall experimental heralding efficiency measured with these detectors of $42.5\%$. By comparing the heralding efficiency of the photon source measured with the camera ($62\%$) we can deduce a lower bound on the iDus 416 camera detection efficiency of $90\pm3\%$. This is in agreement with the manufacturer data that quotes efficiency between $90\%$ and $94\%$ when imaging $800nm$ and cooled to $-35^\circ$C.
The two mono-mode beams imaged on the camera are acquired as fully vertically binned images. The data in grey scale level are then converted into number of photon through the relation $N=S(E_s-E_{\text{Off}})$ where N is the number of photons, $E_s$ is the intensity signal in grey scale level obtained, $E_{\text{Off}}$ is an electronic offset obtained when no light is illuminating the sensor and $S=0.71$ photoelectrons per grey level is the sensitivity of the camera. The photon number values $N_1$ and $N_2$ are then obtained by selecting some region of interest (ROI) delimited over 5 standard deviations of the beam profile, and by summing the intensity of the different pixels inside these ROIs. We used a readout rate of $0.03$ MHz in order to minimise the camera electronic noises, an exposure time of $0.5$ s which lead to a frequency of read around 1 Hz.

\textbf{Acknowledgements:} The authors are grateful to E. Allen, H. V. Cable, C. Erven, A. Laing, E. Lantz, J. Meuller, J. L. O'Brien and M. Thompson for helpful discussions. This work was supported by EPSRC through the QUANTIC hub, ERC, and the Centre for Nanoscience and Quantum Information (NSQI). J.G.R. acknowledges support from an EPSRC established career fellowship. J.C.F.M. acknowledges support from an EPSRC early career fellowship.

\end{document}